\documentclass{article}

\usepackage{PRIMEarxiv}

\usepackage[utf8]{inputenc} 
\usepackage[T1]{fontenc}    
\usepackage{hyperref}       
\usepackage{url}            
\usepackage{booktabs}       
\usepackage{amsfonts}       
\usepackage{nicefrac}       
\usepackage{microtype}      
\usepackage{lipsum}
\usepackage{bm} 
\usepackage{fancyhdr}       
\usepackage{graphicx}       
\usepackage{amsmath}
\graphicspath{{media/}}     

\title{Weak Detonations Revisited:\\Uncovering Its General Nature\\Using Autoignitive Reaction Wave Concept} 

\author{
  Youhi Morii and Kaoru Maruta \\
  Institute of Fluid Science \\ 
  Tohoku University \\
  2-1-1 Katahira, Aoba, Sendai, Miyagi, 980-8577, Japan\\
  \texttt{morii@edyn.ifs.tohoku.ac.jp} \\
}

\begin{document}
\maketitle

\begin{abstract}
Weak detonations have remained experimentally elusive since their theoretical prediction, with previous realization attempts requiring either pathological detonations or Zel'dovich spontaneous waves. Here, we demonstrate that stable weak detonations are naturally achieved through supersonic autoignitive reaction waves—a recently proposed concept describing inherently stable reaction waves determined by inflow velocity conditions and autoignition characteristics. We establish the mathematical equivalence between autoignitive reaction waves and classical Rayleigh flow, proving that supersonic autoignitive reaction waves are indeed weak detonations. The underlying thermodynamic structure reveals Legendre conjugate variables linking normalized enthalpy and fuel consumption. Unlike previous approaches, our universal realization conditions require only that inlet velocity exceed the Chapman-Jouguet velocity and autoignition criteria be met—applicable to any reactive system without specialized chemistry. This framework transforms weak detonations from theoretical curiosities to practically achievable phenomena, opening new possibilities for applications in supersonic combustion systems and astrophysical phenomena.
\end{abstract}

\section{Introduction}\label{sec:introduction}
The theoretical prediction of weak detonations—combustion waves with supersonic flow both upstream and downstream—represents one of the most significant unresolved problems in combustion physics \cite{Lee2008}, with implications spanning from laboratory-scale combustion to astrophysical phenomena including Type Ia supernovae \cite{GarciaSenz1999}.

Classical combustion theory predicts six possible steady one-dimensional reaction wave regimes through Hugoniot-Rayleigh analysis~\cite{Rankine1870,Hugoniot1887,Rayleigh1910}.
Among these, three correspond to supersonic combustion waves (detonations): strong detonation, Chapman-Jouguet detonation, and weak detonation (see Figure \ref{fig.HR})~\cite{Chapman1899,Jouguet1905,Jouguet1906}.
While both strong detonations and Chapman-Jouguet detonations have been extensively observed~\cite{Lee2008}, stable weak detonations remain largely unconfirmed experimentally except in highly specialized chemical systems discussed below.

The stability of weak detonations has been questioned since the foundational work of von Neumann~\cite{vonNeumann1942} and Zel'dovich~\cite{Zeldovich1940}.
Von Neumann showed that shock-initiated weak detonations require "pathological" chemical kinetics with temperature overshoot, experimentally verified only in H$_2$-Cl$_2$ mixtures~\cite{Guenoche1981,Dionne2000}.
Alternatively, Zel'dovich~\cite{Zeldovich1980} proposed the concept of "spontaneous wave" as a shock-free weak detonation mechanism, but its stable propagation requires sequential ignition through precise control of autoignition processes.

Recent theoretical advances have established rigorous connections between zero-dimensional ignition phenomena and one-dimensional reaction waves~\cite{Morii2022,Morii2023}, leading to the discovery of the concept of "autoignitive reaction waves"~\cite{Morii2024}—reaction waves determined by inflow velocity conditions and autoignition characteristics that, under supersonic conditions, maintain flow without shock formation.
Unlike pathological detonations or sequential ignition approaches, autoignitive reaction waves can exist in general chemical systems under appropriate flow conditions.
Here, we resolve the weak detonation stability problem by establishing the mathematical equivalence between autoignitive reaction waves and Rayleigh flow.
This demonstrates that weak detonations can exist as stable phenomena in the form of supersonic autoignitive reaction waves, providing a universal framework applicable to general chemical systems.
Our approach offers a fundamental resolution to this long-standing theoretical challenge while revealing new insights into supersonic combustion mechanisms.

\section{Theoretical framework}\label{sec:theoretical_framework}
\begin{figure}
  \includegraphics[width=\textwidth]{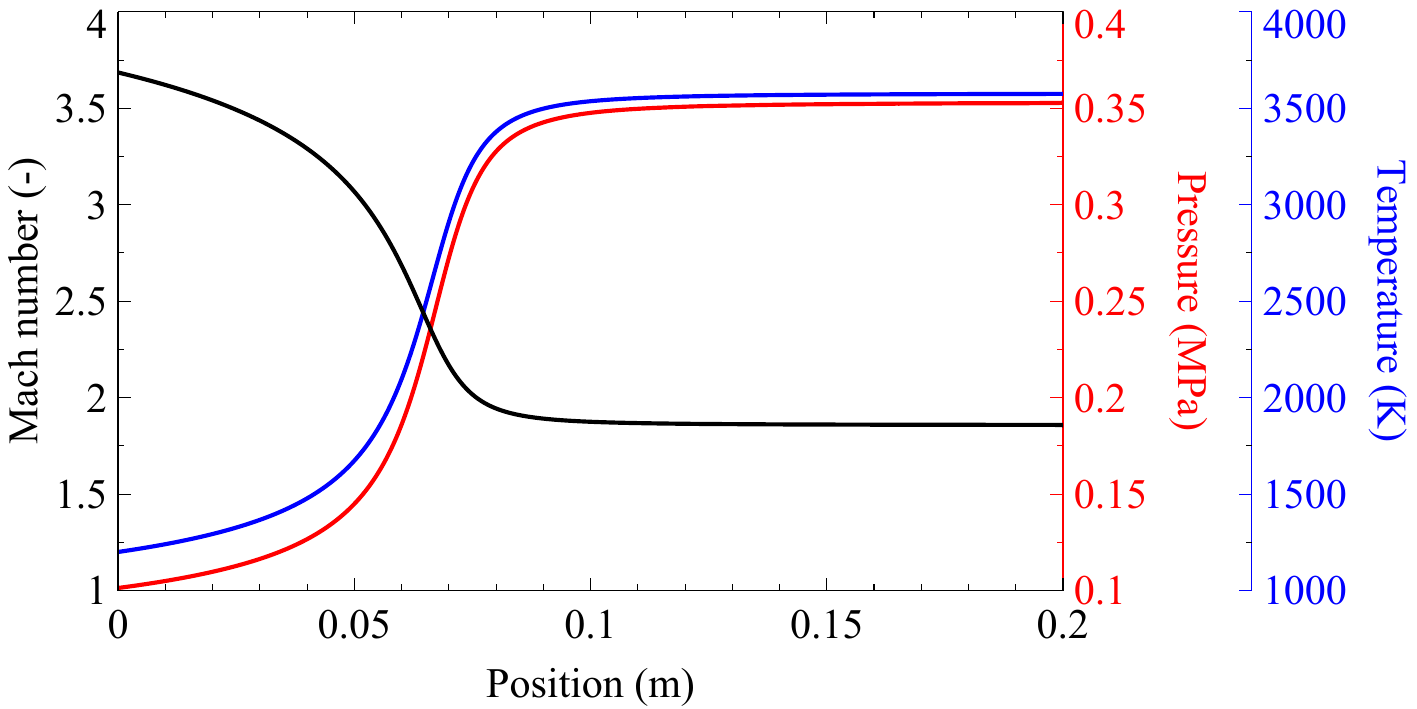}
  \caption{Profiles of supersonic autoignitive reaction wave obtained from unsteady simulations after reaching steady state. The conditions are stoichiometric CH$_4$/Air, inlet pressure of 101325 Pa, inlet temperature of 1200 K, inlet velocity of 2500 m/s, and computational domain of 0.2 m.}
  \label{fig.ARW}
\end{figure}

Our analysis begins with a concrete example of supersonic autoignitive reaction waves. Figure \ref{fig.ARW} presents spatial profiles obtained using the same computational methods as in \cite{Morii2024} but with different flow conditions. This wave exhibits a remarkable property: it maintains supersonic flow both upstream and downstream without any intervening shock structures. These are precisely the characteristics that define weak detonations in classical combustion theory.

The fundamental question we address is: are supersonic autoignitive reaction waves actually weak detonations? To answer this question, we must examine their governing equations. 

\begin{figure}
  \includegraphics[width=\textwidth]{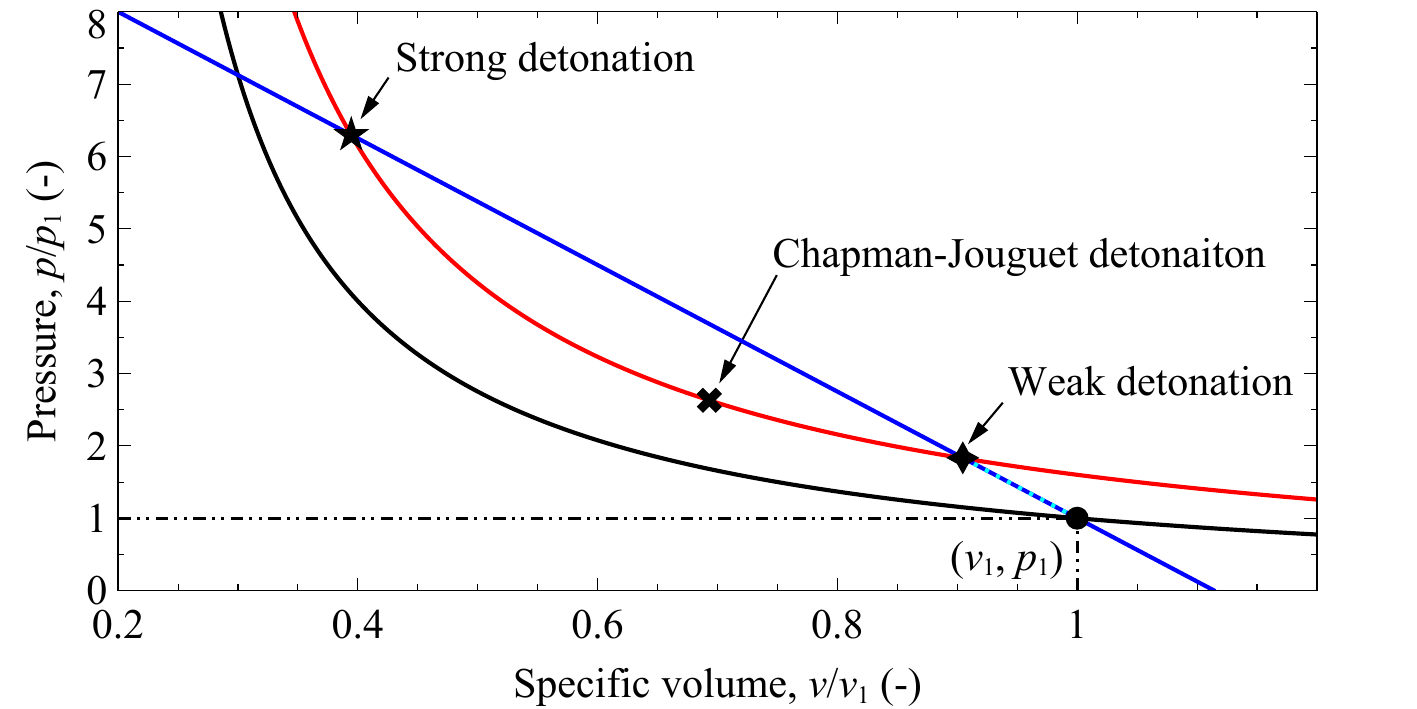}
  \caption{Schematic of Hugoniot-Rayleigh relationship for detonation analysis. The blue line represents the Rayleigh line, the black and red curves show the initial and final Hugoniot curves, respectively. The cyan dotted line represents the autoignitive reaction wave, which this work demonstrates coincides exactly with the Rayleigh line.}
  \label{fig.HR}
\end{figure}

Classical detonation theory relies on the Hugoniot-Rayleigh analysis illustrated in Figure \ref{fig.HR}. In this framework, combustion waves follow specific trajectories in pressure-volume space. The Rayleigh line, shown in blue, represents the locus of states connected by steady flow with heat addition. The key insight of this work is that autoignitive reaction waves, represented by the cyan dotted line in Figure \ref{fig.HR}, precisely coincide with the Rayleigh line. If supersonic autoignitive reaction waves follow these same trajectories, they must be weak detonations.
 
Our approach is therefore to demonstrate that autoignitive reaction waves are mathematically equivalent to Rayleigh flow. We proceed in three steps: First, we review the classical Rayleigh flow equations that govern steady flows with heat addition (Section \ref{subsec:rayleigh_flow}). Second, we derive the complete governing equations for reaction waves and identify the specific conditions—namely, negligible transport effects—under which they simplify (Sections \ref{subsec:1d_reaction_waves} and \ref{subsec:autoignitive_waves}). Finally, we show that under these conditions, the equations become identical to those of Rayleigh flow, establishing that supersonic autoignitive reaction waves are indeed weak detonations.

\subsection{Rayleigh flow}\label{subsec:rayleigh_flow}
\begin{figure}
  \includegraphics[width=\textwidth]{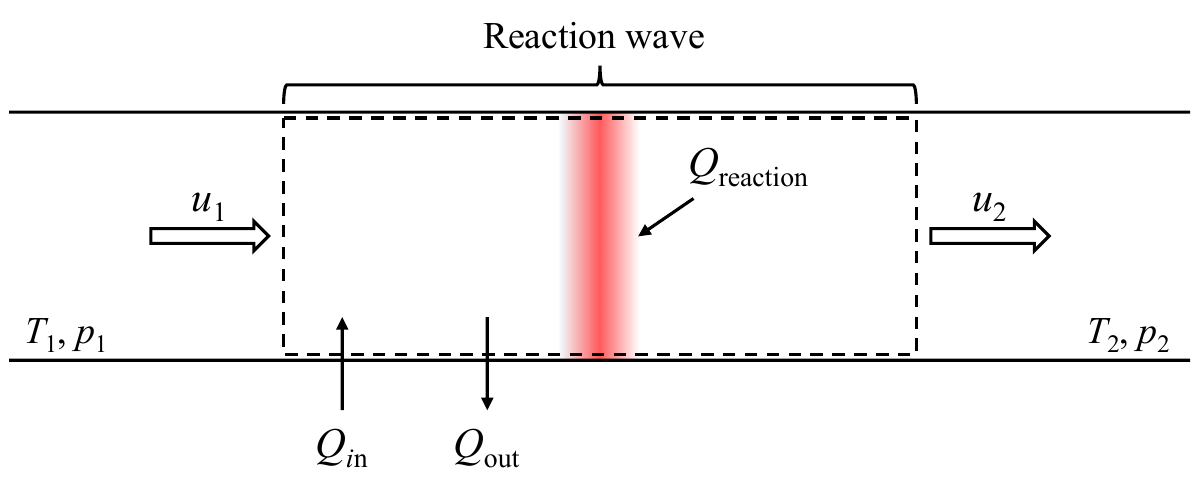}
  \caption{Schematic of Rayleigh flow in a constant area duct. Flow properties change from state 1 (upstream) to state 2 (downstream) through heat addition.}
  \label{fig.RayleighFlow}
\end{figure}

Rayleigh flow describes steady, one-dimensional compressible flow with heat addition in a constant-area duct without friction. This classical model, illustrated in Figure \ref{fig.RayleighFlow}, provides the theoretical foundation for understanding how combustion waves behave thermodynamically.

The governing equations for Rayleigh flow are derived from the fundamental conservation laws:
\begin{equation}
   \frac{\mathrm{d} }{\mathrm{d} x}(\rho u) = 0, 
   \label{eq:ray_cont} 
\end{equation}
\begin{equation}
   \frac{\mathrm{d}}{\mathrm{d} x} (\rho u^2 + p) = 0, 
   \label{eq:ray_mom} 
\end{equation}
\begin{equation}
   \frac{\mathrm{d} }{\mathrm{d} x}\left(c_p T + \frac{u^2}{2}  + q\right) = 0, 
   \label{eq:ray_energy}
\end{equation}
where $\rho$ is the density, $u$ is the velocity, $p$ is the pressure, $c_p$ is the specific heat at constant pressure, $T$ is the temperature, and $q$ represents the cumulative heat added per unit mass up to position $x$. These equations express conservation of mass (\ref{eq:ray_cont}), momentum (\ref{eq:ray_mom}), and energy (\ref{eq:ray_energy}).

The key insight from Rayleigh flow theory is that state changes follow linear trajectories in pressure-specific volume space, known as Rayleigh lines. These trajectories are constrained by momentum conservation: $p - p_1 = \rho_1 u_1^2 (1/v_1 - 1/v)$, where $v = 1/\rho$ is the specific volume and subscript 1 denotes the upstream state. This relationship will prove crucial for identifying weak detonations.

\subsection{One-dimensional reaction waves}\label{subsec:1d_reaction_waves}
We now examine the complete governing equations for reaction waves, including all transport phenomena. Our goal is to identify the specific conditions under which these equations reduce to the Rayleigh flow form. As we will show, this occurs when transport effects become negligible—precisely the condition that defines autoignitive reaction waves.

The general governing equations for one-dimensional reaction waves with transport effects are:
\begin{equation}
   \frac{\partial \rho}{\partial t} + \frac{\partial }{\partial x}(\rho u) = 0,
   \label{eq:continuity}
\end{equation} 
\begin{equation}
  \frac{\partial}{\partial t} (\rho u) + \frac{\partial }{\partial x}\left(\rho u^2 + p -\tau_{xx}\right) = 0, 
  \label{eq:momentum} 
\end{equation} 
\begin{equation}
   \frac{\partial E}{\partial t} + \frac{\partial }{\partial x}\left[(E+p)u + q_x -\tau_{xx}u\right] = -\sum_{s=1}^{N_s}h_s^0\dot{\omega}_s, 
   \label{eq:energy} 
\end{equation} 
\begin{equation} 
  \frac{\partial}{\partial t}(\rho \bm{Y}) + \frac{\partial }{\partial x} \left(\rho \bm{Y} u + \bm{d}_x\right) = \dot{\boldsymbol{\omega}}, 
  \label{eq:species} 
\end{equation}
where $\tau_{xx}$ is the viscous stress, $q_x$ is the heat flux, $\bm{Y}$ is the species mass fraction vector, $\bm{d}_x$ represents species diffusion, $h_s^0$ is the standard enthalpy of formation, and $\dot{\boldsymbol{\omega}}$ denotes species production rates. The total energy $E = \rho e + \rho u^2/2$ includes both internal energy and kinetic energy.

For steady flows, these equations simplify to:
\begin{equation} 
  \frac{\mathrm{d}}{\mathrm{d}x}(\rho u) = 0, 
  \label{eq:continuity-steady} 
\end{equation} 
\begin{equation} 
  \frac{\mathrm{d}}{\mathrm{d}x}\left(\rho u^2 + p -\tau_{xx}\right) = 0, 
  \label{eq:momentum-steady} 
\end{equation} 
\begin{equation} 
  \frac{\mathrm{d}}{\mathrm{d}x}\left[c_p T + \frac{u^2}{2} + \frac{1}{\rho u}\left(q_x-\tau_{xx}u \right)\right]= - \frac{\sum_{s=1}^{N_s}h_s^0\dot{\omega}_s}{\rho u},
  \label{eq:energy_expanded-steady} 
\end{equation} 
\begin{equation} 
  \frac{\mathrm{d}}{\mathrm{d}x} \left(\rho \bm{Y} u + \bm{d}_x\right) = \dot{\boldsymbol{\omega}}.
  \label{eq:species-steady} 
\end{equation}

Comparing these equations with the Rayleigh flow equations (\ref{eq:ray_cont})-(\ref{eq:ray_energy}), we observe additional transport terms: $\tau_{xx}$ in the momentum equation, $q_x$ and viscous work in the energy equation, and $\bm{d}_x$ in the species equation. These transport effects dominate the internal structure of typical reaction waves such as laminar premixed flames, preventing equivalence with Rayleigh flow.

\subsection{Autoignitive reaction waves}\label{subsec:autoignitive_waves}
As shown in the previous section, transport effects prevent general reaction waves from being equivalent to Rayleigh flow. However, a special regime exists where this equivalence can be achieved.

When the inflow velocity exceeds the reaction wave's self-propagation velocity (such as the laminar flame velocity for deflagrations or the Chapman-Jouguet velocity for detonations), the wave loses its self-propagating characteristics and becomes an autoignitive reaction wave~\cite{Morii2024}. In this regime, the wave structure is determined by inflow velocity conditions and autoignition characteristics, with transport effects becoming negligible compared to convective and reactive processes.

This allows us to drop all transport terms from the governing equations, yielding:
\begin{equation}
\frac{\mathrm{d}}{\mathrm{d} x}(\rho u) = 0,
\label{eq:auto_continuity}
\end{equation}
\begin{equation}
\frac{\mathrm{d}}{\mathrm{d} x}(\rho u^2 + p) = 0,
\label{eq:auto_momentum}
\end{equation}
\begin{equation}
  \frac{\mathrm{d}}{\mathrm{d}x}\left(c_p T + \frac{u^2}{2} \right)= -\frac{\sum_{s=1}^{N_s}h_s^0\dot{\omega}_s}{\rho u},
\label{eq:auto_energy}
\end{equation}
\begin{equation}
\frac{\mathrm{d}}{\mathrm{d}x} \left(\rho \bm{Y} u\right) = \dot{\boldsymbol{\omega}}.
\label{eq:auto_species}
\end{equation}
The Eqs. (\ref{eq:auto_continuity}) and (\ref{eq:auto_momentum}) are already equivalent to the Rayleigh flow equations (\ref{eq:ray_cont}) and (\ref{eq:ray_mom}). To establish complete equivalence, we need to demonstrate how Eqs. (\ref{eq:auto_energy}) and (\ref{eq:auto_species}) can be transformed to match Eq. (\ref{eq:ray_energy}).
For clarity, we demonstrate this equivalence using a simplified one-step reaction model.
The extension to complex multi-step reaction mechanisms follows identical mathematical principles (see Supplemental Material [URL will be inserted by publisher]).

The normalized fuel mass fraction $\tilde{Y}_\mathrm{fuel}$ is defined as $\tilde{Y}_\mathrm{fuel} = (Y_{\mathrm{fuel}} - Y_{\mathrm{fuel},1})/(Y_{\mathrm{fuel},0} - Y_{\mathrm{fuel},1})$, where $Y_{\mathrm{fuel},0}$ is the initial fuel mass fraction and $Y_{\mathrm{fuel},1}$ is the final (equilibrium) fuel mass fraction.
Then, $\tilde{Y}_\mathrm{fuel}$ represents the remaining fraction of the initial fuel that has not yet reacted, with $\tilde{Y}_\mathrm{fuel}=1$ in the unburned mixture and $\tilde{Y}_\mathrm{fuel}=0$ in the completely burned mixture.

As a result, Eq. (\ref{eq:auto_species}) can be rewritten as:
\begin{equation}
\frac{\mathrm{d}}{\mathrm{d}x} \left(\rho \tilde{Y}_\mathrm{fuel} u\right) = \tilde{\dot{\omega}}_\mathrm{\,fuel},
\label{eq:fuel_species}
\end{equation}
where $\tilde{\dot{\omega}}_\mathrm{\,fuel} = \dot{\omega}_{\mathrm{\,fuel}}/(Y_{\mathrm{fuel},0} - Y_{\mathrm{fuel},1})$ represents the normalized reaction rate of the fuel.
Using the product rule and the continuity equation (\ref{eq:auto_continuity}), Eq. (\ref{eq:fuel_species}) reduces to:
\begin{equation}
\frac{\mathrm{d}\tilde{Y}_\mathrm{fuel}}{\mathrm{d}x} = \frac{\tilde{\dot{\omega}}_\mathrm{\,fuel}}{\rho u}.
\label{eq:dYf_dx}
\end{equation}
This equation establishes the crucial relationship between the fuel mass fraction gradient and reaction rate.
Using this equation, we need not solve the other species conservation equations because once $\tilde{Y}_\mathrm{fuel}$ is determined, all other species mass fractions $\bm{Y}$ can be obtained.

Using one-step chemical reactions, the heat release is directly proportional to the fuel consumption rate.
We define $Q_\mathrm{reaction}$ as the total heat released per unit mass of mixture when the fuel is completely consumed from its initial to final state, giving us:
\begin{equation}
  \sum_{s=1}^{N_s}h_s^0\dot{\omega}_s = Q_\mathrm{reaction} \tilde{\dot{\omega}}_\mathrm{\,fuel}.
\end{equation}
This allows us to substitute the chemical source term in the energy equation with the fuel consumption rate:
\begin{equation}
  \frac{\mathrm{d}}{\mathrm{d}x}\left(c_p T + \frac{u^2}{2} \right)= -\frac{Q_\mathrm{reaction} \tilde{\dot{\omega}}_\mathrm{\,fuel}}{\rho u}.
\end{equation}
Using the relationship from Eq. (\ref{eq:dYf_dx}), we can eliminate $\tilde{\dot{\omega}}_\mathrm{\,fuel}/(\rho u)$ and express the energy equation directly in terms of the fuel mass fraction gradient:
\begin{equation}
  \frac{\mathrm{d}}{\mathrm{d}x}\left(c_p T + \frac{u^2}{2} \right)= -Q_\mathrm{reaction}\frac{\mathrm{d}\tilde{Y}_\mathrm{fuel}}{\mathrm{d}x}.
\end{equation}
Rearranging the terms to isolate the derivatives:
\begin{equation}
  \frac{\mathrm{d}}{\mathrm{d}x}\left(\frac{c_p T + u^2/2 }{Q_\mathrm{reaction}}\right)= -\frac{\mathrm{d}\tilde{Y}_\mathrm{fuel}}{\mathrm{d}x}.
  \label{eq:legendre}
\end{equation}
This reveals that the normalized enthalpy $\tilde{H}(x) = (c_p T + u^2/2)/Q_\mathrm{reaction}$ and $\tilde{Y}_\mathrm{fuel}(x)$ are Legendre conjugate variables, providing the thermodynamic foundation for weak detonation stability (see Supplemental Material [URL will be inserted by publisher]).

Rearranging Eq. (\ref{eq:legendre}), we obtain:
\begin{equation}
  \frac{\mathrm{d}}{\mathrm{d}x}\left(c_p T + \frac{u^2}{2} + Q_\mathrm{reaction}\tilde{Y}_\mathrm{fuel} \right)= 0.
\end{equation}

The governing equations for autoignitive reaction waves can therefore be summarized in a form identical to Rayleigh flow:
\begin{equation}
\frac{\mathrm{d}}{\mathrm{d} x}(\rho u) = 0,
\end{equation}
\begin{equation}
\frac{\mathrm{d}}{\mathrm{d} x}(\rho u^2 + p) = 0,
\end{equation}
\begin{equation}
\frac{\mathrm{d}}{\mathrm{d}x}\left(c_p T + \frac{u^2}{2} + q_\mathrm{total}\right) = 0
\end{equation}
where $q_\mathrm{total}(x) =Q_\mathrm{reaction}\tilde{Y}_\mathrm{fuel} + q_\mathrm{wall}$, with $q_\mathrm{wall}$ representing the cumulative heat transfer from the walls per unit mass.
This is exactly analogous to the heat addition term $q$ in the classical Rayleigh flow equation (\ref{eq:ray_energy}). We have thus established that autoignitive reaction waves are governed by the same equations as Rayleigh flow. Under supersonic conditions, these waves therefore represent weak detonations.

\subsection{Physical realization of weak detonations}\label{subsec:physical_realization}
Having established the mathematical equivalence between autoignitive reaction waves and Rayleigh flow, we now identify the physical conditions required to achieve weak detonations in practice:

\begin{enumerate}
\item \textbf{Fluid-dynamic requirement}: The inlet velocity must exceed the Chapman-Jouguet detonation velocity:
\begin{equation}
u_1 > D_\mathrm{CJ}
\end{equation}

\item \textbf{Chemical-kinetic requirement}: The temperature and residence time must allow autoignition to occur:
\begin{equation}
L \gg  u_1\tau_{\text{ignition}}
\end{equation}
\end{enumerate}
where $u_1$ is the inlet flow velocity, $D_\mathrm{CJ}$ is the Chapman-Jouguet detonation velocity, $L$ is the computational or physical domain length, and $\tau_{\text{ignition}}$ is the ignition delay time.

To illustrate these conditions, consider the example shown in Figure \ref{fig.ARW}. For the stoichiometric CH$_4$/Air mixture at 1200 K and 101325 Pa, the Chapman-Jouguet detonation velocity is $D_{\mathrm{CJ}} = 1978.12$ m/s and the ignition delay time is $\tau_{\text{ignition}} = 3.85 \times 10^{-5}$ s. With an inlet velocity of $u_1 = 2500$ m/s, both conditions are satisfied: (1) $u_1 > D_{\mathrm{CJ}}$ since $2500 > 1978.12$ m/s, and (2) $L \gg u_1\tau_{\text{ignition}}$ since $u_1 \tau_{\text{ignition}} = 0.096$ m is much smaller than the computational domain length $L = 0.2$ m. This ensures that autoignition occurs well within the domain, enabling the formation of a stable weak detonation.

These conditions differ fundamentally from previous approaches: no pathological chemistry or external energy deposition is required. Instead, weak detonations can be achieved through fluid-dynamic control alone.

\section{Conclusions}\label{sec:conclusions}

We have demonstrated that autoignitive reaction waves are mathematically equivalent to Rayleigh flow, establishing that weak detonations can exist as stable supersonic phenomena. This resolves the long-standing problem of why weak detonations, though theoretically predicted, have remained experimentally elusive.

Our analysis reveals three key insights: (1) transport effects must be negligible for equivalence with Rayleigh flow, which occurs in autoignitive reaction waves; (2) the Legendre conjugate relationship between enthalpy and fuel consumption provides the thermodynamic foundation for this equivalence; and (3) weak detonations require only that inlet velocity exceed the CJ velocity and that autoignition conditions be met—no specialized chemistry or external energy sources are needed.

This framework transforms weak detonations from theoretical curiosities to practically achievable phenomena, with broad applications in supersonic combustion systems and astrophysical phenomena.

\section*{Acknowledgments}
This work was partially supported by JSPS KAKENHI Grant Number 19KK0097.

\bibliographystyle{unsrt}  
\bibliography{library}

\clearpage

\section*{Supplemental material}\label{sec:supplemental}

\subsection*{S1. Extension to complex chemical mechanisms}\label{subsec:extension_complex}

While our main text demonstrated the equivalence between autoignitive reaction waves and Rayleigh flow using a one-step reaction model, real combustion systems involve multiple species and reactions. Here, we extend our theoretical framework to complex chemical mechanisms.

For complex reaction mechanisms, we define a reaction progress variable $C$ as:
\begin{equation}
C = F(\bm{Y}) = F(Y_1, Y_2, ..., Y_{N_s}),
\end{equation}
where $F$ is a function that increases monotonically from 0 (unreacted) to 1 (fully reacted). This can be defined based on reactant consumption, product formation, or heat release, depending on which best characterizes the specific chemical system under consideration.

With a properly defined progress variable, the heat release term can be expressed as:
\begin{equation}
\sum_{s=1}^{N_s}h_s^0\dot{\omega}_s = -Q_\mathrm{total}\dot{\omega}_{\,C}.
\end{equation}
where $Q_\mathrm{total}$ is the total heat release per unit mass for complete reaction and $\dot{\omega}_{\,C} = \mathrm{d}C/\mathrm{d}t$ is the reaction rate of the progress variable.

From the energy and species conservation equations, we derive:
\begin{equation}
\frac{\mathrm{d}}{\mathrm{d}x}\left(\frac{c_p T + u^2/2}{Q_\mathrm{total}}\right) = \frac{\mathrm{d}C}{\mathrm{d}x}.
\end{equation}
Integrating, we obtain:
\begin{equation}
\frac{\mathrm{d}}{\mathrm{d}x}\left(c_p T + \frac{u^2}{2} - Q_\mathrm{total}C\right) = 0.
\end{equation}

This final form is identical to the Rayleigh flow energy equation, with $Q_\mathrm{total}C$ serving as the cumulative heat release term.
This demonstrates that the mathematical equivalence between autoignitive reaction waves and Rayleigh flow holds regardless of the chemical mechanism's complexity.
The equivalence is valid provided that the reaction follows a unique path in composition space, transport effects are negligible, and the progress variable changes monotonically throughout the reaction zone.

\subsection*{S2. Legendre transformation in reaction wave analysis}\label{subsec:legendre_transformation}

The key mathematical structure underlying our analysis is the Legendre conjugate relationship between the normalized enthalpy $\tilde{H}(x) = (c_p T + u^2/2)/Q_\mathrm{reaction}$ and the reaction progress $\tilde{Y}_\mathrm{fuel}(x)$:
\begin{equation}
\frac{\mathrm{d}\tilde{H}}{\mathrm{d}x} = -\frac{\mathrm{d}\tilde{Y}_\mathrm{fuel}}{\mathrm{d}x}.
\end{equation}

This duality has three important implications:

\textbf{Physical interpretation}: The relationship shows that energy accumulation and chemical transformation are fundamentally coupled. Energy gained by the system through heating precisely matches the energy released by chemical reaction, establishing a direct link between thermodynamic and chemical states.

\textbf{Mathematical simplification}: The Legendre structure allows us to eliminate the species conservation equation by expressing chemical progress in terms of enthalpy changes. This reduction from a system of coupled PDEs to a single energy equation is what enables the equivalence with Rayleigh flow.

\textbf{Generality}: This structure holds for any reaction mechanism (not just one-step reactions) when expressed in terms of an appropriate progress variable $C$, making our results applicable to realistic combustion systems.

\end{document}